\documentclass[preprint,12pt]{elsarticle}

\usepackage{amsmath,amssymb,bm}
\usepackage{booktabs}
\usepackage{graphicx}
\usepackage{microtype}
\usepackage[hidelinks]{hyperref}

\biboptions{sort&compress}

\hypersetup{
  pdftitle={Entropy Production and Thermalization in Heavy-Quark Langevin Dynamics},
  pdfauthor={A. M. Silva}
}

\newcommand{\dd}{\mathrm{d}}
\newcommand{\ee}{\mathrm{e}}
\newcommand{\avg}[1]{\left\langle #1\right\rangle}
\newcommand{\vect}[1]{\bm{#1}}

\newcommand{\KL}{D_{\mathrm{KL}}}

\journal{Physica A}

\begin{document}

\begin{frontmatter}

\title{Entropy Production and Thermalization in\\
Heavy-Quark Langevin Dynamics}

\author{A. M. Silva}

\renewcommand{\theaffn}{}
\affiliation{organization={Instituto de F\'isica, Universidade de S\~ao Paulo},
  addressline={Rua do Mat\~ao, 1371, Cidade Universit\'aria},
  postcode={05508-090},
  city={S\~ao Paulo},
  state={SP},
  country={Brazil}}
\makeatletter
\g@addto@macro\elsaddress{\vskip0.3em{\upshape\ttfamily alisson.matheus.silva@usp.br}\par}
\makeatother

\begin{abstract}
Heavy quarks are often treated as Brownian probes of the quark-gluon plasma.
We formulate the irreversible entropy balance of their momentum distribution
and separate the production generated by collisions from the driving caused by
an evolving medium.  For linear dynamics the production splits into two positive
contributions, from directed momentum and from the momentum width.  A static
bath obeys an exact $H$-theorem.  A cooling bath adds an explicit driving term,
and longitudinal Bjorken expansion enters as a
separate kinematic drive rather than as collisional production.  For a Bjorken
temperature protocol we derive closed-form strong- and weak-coupling limits of
the total production.  At strong coupling it exceeds the initial relative
entropy by a term linear in the relaxation time, whose quasistatic part has the
thermodynamic-length form.  At weak coupling it vanishes inversely with the
relaxation time.  A positive strong-coupling slope therefore guarantees maximal
dissipation at intermediate coupling.  The same structure holds for a
relativistic Fokker--Planck equation with a J\"uttner stationary state, solved
numerically for charm and bottom quarks under the temperature protocol.
Linear-response coefficients computed from the static generator reproduce both
limits, and the numerical entropy balance is satisfied to high accuracy.
\end{abstract}


\end{frontmatter}

\section{Introduction}
\label{sec:introduction}

Heavy quarks are created in the first hard scatterings of a relativistic nuclear
collision.  Because their masses are large compared with the temperatures
reached by the plasma, they are not produced thermally in appreciable numbers
and remain identifiable probes throughout the evolution.  If each collision
transfers only a small fraction of a heavy quark's momentum, the cumulative
effect of many collisions is described naturally by a Fokker--Planck or
Langevin equation \cite{Svetitsky1988,Moore2005,vanHees2005}.  That picture has
grown into a central part of heavy-flavor phenomenology, including relativistic
transport coupled to hydrodynamic backgrounds and realistic hadronization
\cite{Akamatsu2009,Beraudo2009,HeFriesRapp2012,Cao2013,Ke2018,He2023}.
Comparisons with the full Boltzmann equation also make its domain of validity
clear: the approximation is strongest for bottom quarks and for dynamics
dominated by soft momentum transfer, while charm at high temperature can retain
visible non-Brownian corrections \cite{Das2014}.

From the standpoint of statistical mechanics, a heavy quark is therefore a
Brownian particle in a bath that cools on a time scale comparable to its own
relaxation time.  The Langevin equation contains more information than that
relaxation time.  It also defines a probability current and therefore an
entropy balance.  In stochastic thermodynamics, the change of distribution
entropy is separated into irreversible production inside the reduced system and
entropy exchanged with its environment \cite{Sekimoto2010,Seifert2012}.  Landi,
Tom\'e, and de Oliveira showed that this decomposition can be written exactly in
terms of means and covariances for linear Langevin systems, including variables
with odd time-reversal parity \cite{Landi2013}.  The role of odd variables such
as momentum in full phase space was developed systematically by Spinney and
Ford \cite{SpinneyFord2012PRL,SpinneyFord2012}.  Driven Brownian systems also
make clear that motion of the reference state contributes separately to a
relative-entropy balance \cite{SchmiedlSeifert2008,FordLakerCharlesworth2015}.
For slowly driven systems, finite-time thermodynamics further predicts an
excess dissipation proportional to the relaxation time, organized by a
thermodynamic metric \cite{SalamonBerry1983,SivakCrooks2012}.

We specialize that structure to heavy-quark diffusion.  The Gaussian formula
derived below is a direct heavy-quark instance of the general linear result of
Landi et al.\ \cite{Landi2013}, rather than a new entropy-production law.  Its
value is that it exposes a clean diagnostic for transport calculations: one
positive term measures the loss of directed momentum and another measures the
mismatch between the momentum width and its thermal value.  The new results
concern the driven problem.  For the Bjorken temperature protocol we obtain the
total production in closed form at strong and weak coupling.  The strong-coupling
correction contains, besides the thermodynamic-length term, a contribution from
the initial relaxation taking place against a moving target.  Together the two
limits prove that the dissipation is maximal at intermediate coupling whenever
the strong-coupling slope is positive.  We then solve a relativistic
detailed-balance generator for charm and bottom, scan the production across the
diffusion strength, and compute the same two limits from the static generator by
linear response.

The distinction from existing entropy studies in the QGP is also important.
Open-quantum-system treatments have followed the entropy and free energy of
heavy quarks or quarkonia through reduced density matrices
\cite{BlaizotEscobedo2018}.  Here the object is instead the classical
single-particle momentum density generated by a Langevin transport equation.
The entropy studied here belongs to that reduced distribution.  It quantifies
how much information about the initial momentum survives and how irreversibly
the distribution approaches its local thermal target.  It is not the total
thermodynamic entropy of the QGP.  Because the two notions are kept separate,
the quantity derived below can be evaluated along any transport trajectory
without being confused with the viscous entropy production of the bulk medium.

\section{Heavy-quark Ornstein--Uhlenbeck process}
\label{sec:ou}

We first consider a homogeneous medium in its local rest frame and an isotropic
$d$-dimensional momentum process,
\begin{equation}
  \dd\vect p
  =-\gamma(t)\vect p\,\dd t
  +\sqrt{2D(t)}\,\dd\vect W_t,
  \label{eq:langevin}
\end{equation}
where $\gamma(t)>0$ is the drag coefficient and $D(t)>0$ is the momentum-space
diffusion coefficient per Cartesian direction.  The Wiener increments obey
\begin{equation}
  \avg{\dd W_i(t)\dd W_j(t)}=\delta_{ij}\dd t.
\end{equation}
Equivalently, if the random force is written as $\vect\xi(t)$, the convention is
\begin{equation}
  \avg{\xi_i(t)\xi_j(t')}=\kappa(t)\delta_{ij}\delta(t-t'),
  \qquad \kappa(t)=2D(t).
  \label{eq:kappa}
\end{equation}
Keeping $\kappa$ or $D$ explicit is necessary for dimensional consistency,
since a choice $D=1/2$ is meaningful only after a complete
nondimensionalization.

The probability density $f(\vect p,t)$ is normalized with respect to
$\dd^d p$.  The standard Langevin--Fokker--Planck correspondence
\cite{Risken1989} gives
\begin{equation}
  \partial_t f
  =\vect\nabla_p\cdot
  \left[\gamma(t)\vect p f+D(t)\vect\nabla_p f\right]
  =-\vect\nabla_p\cdot\vect J,
  \label{eq:fokker-planck}
\end{equation}
with probability current
\begin{equation}
  \vect J=-\gamma\vect p f-D\vect\nabla_p f.
  \label{eq:current}
\end{equation}
For a Gaussian initial state with mean $\vect m_0$ and isotropic covariance
$\theta_0 I_d$,
\begin{equation}
  f(\vect p,0)=\frac{1}{(2\pi\theta_0)^{d/2}}
  \exp\left[-\frac{|\vect p-\vect m_0|^2}{2\theta_0}\right],
  \qquad \theta_0>0,
  \label{eq:initial}
\end{equation}
the state remains Gaussian,
\begin{equation}
  f(\vect p,t)=\frac{1}{[2\pi\theta(t)]^{d/2}}
  \exp\left[-\frac{|\vect p-\vect m(t)|^2}{2\theta(t)}\right].
  \label{eq:gaussian}
\end{equation}
Its moments obey
\begin{subequations}
\begin{align}
  \dot{\vect m}&=-\gamma\vect m,
  \label{eq:mean-ode}\\
  \dot\theta&=-2\gamma\theta+2D.
  \label{eq:variance-ode}
\end{align}
\end{subequations}
Defining
\begin{equation}
  G(t)=\exp\left[-\int_0^t\gamma(s)\,\dd s\right],
\end{equation}
their solutions are
\begin{subequations}
\begin{align}
  \vect m(t)&=G(t)\vect m_0,
  \label{eq:mean-solution}\\
  \theta(t)&=G^2(t)
  \left[\theta_0+2\int_0^t\frac{D(s)}{G^2(s)}\,\dd s\right].
  \label{eq:variance-solution}
\end{align}
\end{subequations}

\section{Entropy balance and positive decomposition}
\label{sec:entropy}

The differential entropy of the heavy-quark momentum distribution is
\begin{equation}
  S(t)=-\int f(\vect p,t)\ln f(\vect p,t)\,\dd^d p.
  \label{eq:entropy-definition}
\end{equation}
An implicit fixed momentum scale can be inserted in the logarithm.  It changes
$S$ by an additive constant but does not affect any rate below.  For
Eq.~\eqref{eq:gaussian},
\begin{equation}
  S(t)=\frac d2\left[1+\ln(2\pi\theta)\right],
  \qquad
  \dot S=d\left(\frac{D}{\theta}-\gamma\right).
  \label{eq:entropy-rate}
\end{equation}

All momentum components are odd under time reversal.  With no external force,
the current in Eq.~\eqref{eq:current} is therefore the irreversible current.
The standard current expression for the entropy-production rate is
\cite{Landi2013}
\begin{equation}
  \Pi(t)=\int\frac{\vect J^{\mathsf T}(D I_d)^{-1}\vect J}{f}\,\dd^d p
  \geq 0.
  \label{eq:production-current}
\end{equation}
Using
\begin{equation}
  \vect\nabla_p f=-\frac{\vect p-\vect m}{\theta}f
\end{equation}
and carrying out the Gaussian average gives
\begin{equation}
  \boxed{
  \Pi(t)=
  d\frac{[D(t)-\gamma(t)\theta(t)]^2}{D(t)\theta(t)}
  +\frac{\gamma^2(t)|\vect m(t)|^2}{D(t)}.}
  \label{eq:production-main}
\end{equation}
Equation~\eqref{eq:production-main} is the isotropic Gaussian specialization
of the linear-Langevin result in Ref.~\cite{Landi2013}.  It makes positivity
explicit and separates two mechanisms,
\begin{equation}
  \Pi=\Pi_{\mathrm{width}}+\Pi_{\mathrm{drift}}.
\end{equation}
The first term vanishes when the variance equals its instantaneous fixed-point
value $D/\gamma$, and the second vanishes when the mean momentum is zero.

Writing the entropy balance as
\begin{equation}
  \dot S=\Pi-\Phi,
  \label{eq:entropy-balance}
\end{equation}
where $\Phi$ is positive for entropy flow from the heavy quark to the bath, one
finds
\begin{equation}
  \boxed{
  \Phi(t)=
  \frac{\gamma^2(t)}{D(t)}
  \left[|\vect m(t)|^2+d\theta(t)\right]-d\gamma(t).}
  \label{eq:flux-general}
\end{equation}
Unlike $\Pi$, the flux may have either sign during a transient.  Equations
\eqref{eq:entropy-rate}, \eqref{eq:production-main}, and
\eqref{eq:flux-general} satisfy Eq.~\eqref{eq:entropy-balance} identically.

\section{Static thermal bath and exact \texorpdfstring{$H$}{H}-theorem}
\label{sec:static}

For a nonrelativistic heavy quark of mass $M$ in a static bath at temperature
$T$, detailed balance requires the Einstein relation
\begin{equation}
  D=\gamma MT,
  \qquad \kappa=2\gamma MT.
  \label{eq:einstein}
\end{equation}
The stationary density is the Maxwell distribution with covariance $MT I_d$.
The entropy balance reduces to
\begin{subequations}
\begin{align}
  \Pi&=\gamma\left[
  \frac{|\vect m|^2}{MT}
  +d\frac{(\theta-MT)^2}{MT\theta}\right],
  \label{eq:production-thermal}\\
  \dot S&=d\gamma\left(\frac{MT}{\theta}-1\right),
  \label{eq:entropy-rate-thermal}\\
  \Phi&=\gamma\left[
  \frac{|\vect m|^2+d\theta}{MT}-d\right].
  \label{eq:flux-thermal}
\end{align}
\end{subequations}
The mean kinetic energy is
$\avg{E_k}=(|\vect m|^2+d\theta)/(2M)$.  It follows directly from the moment
equations that $\Phi$ equals the mean heat flow from the reduced heavy-quark
system to the bath divided by $T$.

The structure is especially transparent in terms of
\begin{equation}
  x=\frac{\theta}{MT},
  \qquad
  r^2=\frac{|\vect m|^2}{MT},
  \qquad
  \tau=\gamma t.
\end{equation}
For constant coefficients,
\begin{subequations}
\begin{align}
  x(\tau)&=1+(x_0-1)\ee^{-2\tau},\\
  r^2(\tau)&=r_0^2\ee^{-2\tau},
\end{align}
\end{subequations}
and
\begin{subequations}
\begin{align}
  \frac{\Pi}{\gamma}&=r^2+d\frac{(x-1)^2}{x},
  \label{eq:scaled-production}\\
  \frac{\dot S}{\gamma}&=d\left(\frac1x-1\right),\\
  \frac{\Phi}{\gamma}&=r^2+dx-d.
\end{align}
\end{subequations}

Let $f_{\mathrm{eq}}$ denote the Maxwell density.  The relative entropy is
\begin{equation}
  \mathcal H(t)=\KL(f\Vert f_{\mathrm{eq}})
  =\frac12\left[r^2+d(x-1-\ln x)\right].
  \label{eq:relative-entropy-static}
\end{equation}
Differentiation gives the exact $H$-theorem
\begin{equation}
  \boxed{\dot{\mathcal H}=-\Pi.}
  \label{eq:h-theorem}
\end{equation}
Consequently, the integrated entropy production is
\begin{equation}
  \Sigma(t)=\int_0^t\Pi(s)\,\dd s
  =\mathcal H(0)-\mathcal H(t),
  \label{eq:integrated-static}
\end{equation}
and its asymptotic value is
\begin{equation}
  \boxed{
  \Sigma_\infty=\frac12\left[
  r_0^2+d(x_0-1-\ln x_0)\right].}
  \label{eq:total-static}
\end{equation}
This quantity depends on the initial distance from equilibrium, but not on the
relaxation time.  The latter only sets how rapidly the fixed total is produced.

\begin{figure}[t]
  \centering
  \includegraphics[width=\linewidth]{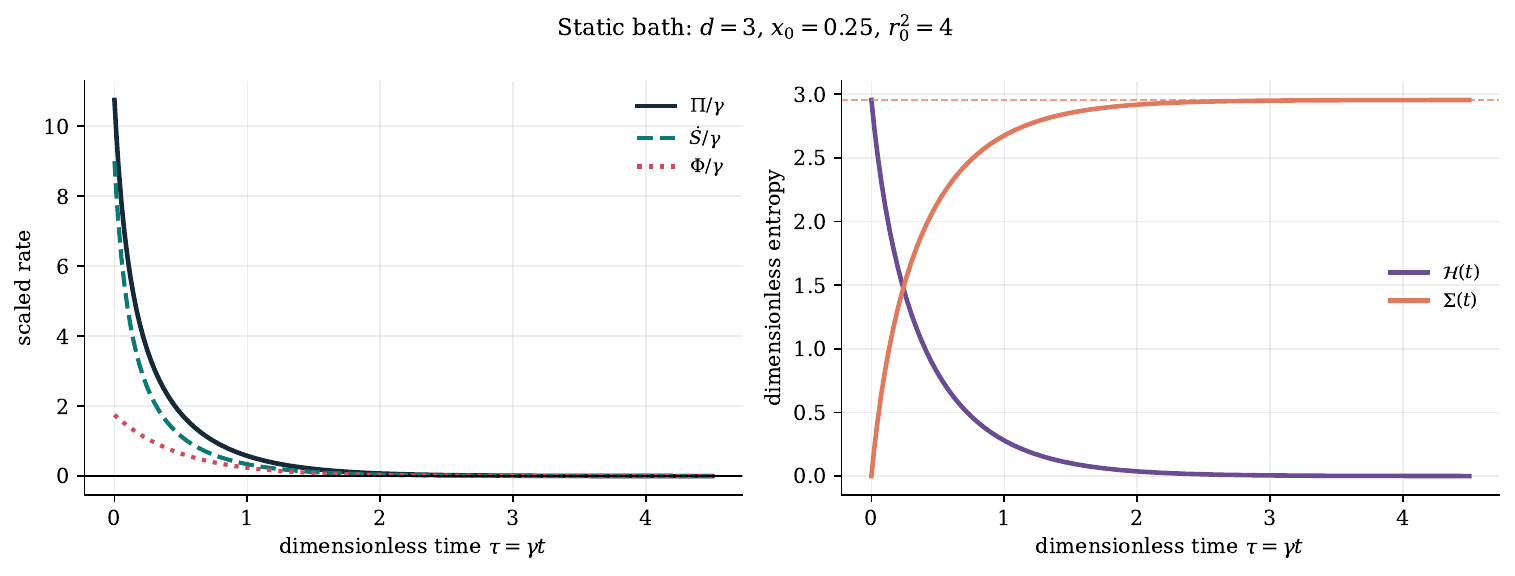}
  \caption{Static-bath entropy balance for $d=3$, initial scaled variance
  $x_0=0.25$, and scaled mean momentum $r_0^2=4$.  Left: entropy production,
  system entropy rate, and entropy flux.  Right: relative entropy and integrated
  entropy production.  Equation~\eqref{eq:h-theorem} implies that the two curves
  on the right sum to the constant $\mathcal H(0)$.}
  \label{fig:static}
\end{figure}

Figure~\ref{fig:static} illustrates the balance for $d=3$, $x_0=0.25$, and
$r_0^2=4$.  The production is non-negative, while the relative entropy decays
by precisely the accumulated production.  The formal limit $x_0\to0$,
corresponding to a deterministic initial momentum, gives
$\Sigma_\infty\sim-(d/2)\ln x_0\to\infty$.  Thus the commonly used
delta-function idealization is outside the differential-entropy calculation.
A finite-width approximation is required whenever this entropy budget is to be
evaluated.  This observation does not restrict simulations that sample a
resolved initial spectrum.

\section{Driven Gaussian benchmark}
\label{sec:cooling}

We next allow the reference bath to cool.  The purpose of this section is
analytic: the Gaussian process remains exactly closed and displays how a moving
thermal target modifies the static $H$-theorem.  No physical charm or bottom
spectrum is assigned to the nonrelativistic solution.  For later comparison we
use the temperature history of ideal boost-invariant expansion
\cite{Bjorken1983},
\begin{equation}
  T(t)=T_0\left(\frac{\tau_0}{\tau_0+t}\right)^{1/3},
  \label{eq:bjorken-temperature}
\end{equation}
where $t=0$ denotes the initial proper time $\tau_0$.  The spatial diffusion
coefficient is parameterized by the dimensionless quantity
\begin{equation}
  c_D=2\pi T D_s.
\end{equation}
Within the nonrelativistic model, $D_s=T/(M\gamma)$ gives
\begin{equation}
  \gamma_Q(t)=\frac{2\pi T^2(t)}{c_D M_Q},
  \qquad D_Q(t)=\gamma_Q(t)M_QT(t).
  \label{eq:cooling-coefficients}
\end{equation}
The corresponding mean propagator is
\begin{equation}
  G_Q(t)=\exp\left\{-3\gamma_Q(0)\tau_0
  \left[\left(1+\frac{t}{\tau_0}\right)^{1/3}-1\right]\right\}.
  \label{eq:cooling-propagator}
\end{equation}
Natural units are understood.  The variance follows from
Eq.~\eqref{eq:variance-solution} without a quasistatic approximation.

For a time-dependent temperature, define the instantaneous reference density
$f_{\mathrm{eq}}(\vect p,t)$ with covariance $M T(t) I_d$ and
\begin{equation}
  \mathcal H(t)=\KL[f(t)\Vert f_{\mathrm{eq}}(t)].
\end{equation}
The production retains the positive instantaneous form
\begin{equation}
  \Pi(t)=\gamma(t)\left[
  \frac{|\vect m(t)|^2}{MT(t)}
  +d\frac{[\theta(t)-MT(t)]^2}{MT(t)\theta(t)}
  \right].
  \label{eq:cooling-production}
\end{equation}
However, the static $H$-theorem acquires a driving contribution,
\begin{equation}
  \boxed{
  \dot{\mathcal H}
  =-\Pi
  +\frac{\dot T}{2T}
  \left[d-\frac{|\vect m|^2+d\theta}{MT}\right].}
  \label{eq:driven-h-theorem}
\end{equation}
The second term is the change in relative entropy caused by motion of the
thermal reference state itself.  Hence $\int\Pi\,\dd t$ need not equal the
initial relative entropy and may depend nontrivially on the relaxation rate.

Equation~\eqref{eq:driven-h-theorem} is already enough to show why production
cannot be inferred from the change of relative entropy alone.  It is also the
nonrelativistic limit of the relativistic identity tested numerically below.

\subsection{Strong- and weak-coupling limits}
\label{sec:limits}

The driving term in Eq.~\eqref{eq:driven-h-theorem} depends on the state only
through the scaled excess kinetic energy
\begin{equation}
  u(t)=\frac{|\vect m|^2+d\theta}{MT}-d.
\end{equation}
Integrating Eq.~\eqref{eq:driven-h-theorem} up to the final time $t_f$ gives
\begin{equation}
  \Sigma(t_f)=\mathcal H(0)-\mathcal H(t_f)
  -\int_0^{t_f}\frac{\dot T}{2T}\,u\,\dd t,
  \label{eq:sigma-u}
\end{equation}
and the moment equations close on $u$ alone,
\begin{equation}
  \dot u=-\left(2\gamma+\frac{\dot T}{T}\right)u-d\,\frac{\dot T}{T}.
  \label{eq:u-ode}
\end{equation}
At strong coupling, $\gamma\tau\gg1$ for the cooling time scale $\tau$,
Eq.~\eqref{eq:u-ode} has the uniformly valid solution
$u\simeq u_0G^2+d|\dot T|/(2\gamma T)$: an initial transient plus a
quasistatic lag behind the cooling target.  Because $\mathcal H(t_f)$ is
quadratic in the lag, it is of order $\gamma^{-2}$, and
Eq.~\eqref{eq:sigma-u} yields, for any smooth cooling protocol,
\begin{equation}
  \Sigma(t_f)\simeq\mathcal H(0)
  +\frac{u_0|\dot T(0)|}{4\gamma(0)T_0}
  +\frac d4\int_0^{t_f}\left(\frac{\dot T}{T}\right)^2\frac{\dd t}{\gamma}
  +O(\gamma^{-2}).
  \label{eq:strong-general}
\end{equation}
The last term is the quasistatic excess dissipation of finite-time
thermodynamics.  Written as $\int\dot T\,\zeta(T)\,\dot T\,\dd t$, it defines
the friction coefficient $\zeta=d/(4\gamma T^2)$, proportional to the
relaxation time, which is the metric of the thermodynamic-length description of
slowly driven systems \cite{SalamonBerry1983,SivakCrooks2012}.  The middle term
has no counterpart in that description.  It is the cost of performing the
initial relaxation against a target that moves while the relaxation is under
way.

For the Bjorken protocol, $\dot T/T=-1/[3(\tau_0+t)]$ and
$\gamma\propto T^2$.  Defining the relaxation-time ratio
\begin{equation}
  \epsilon=\frac{1}{\gamma(0)\tau_0}
  =\frac{c_D M\hbar c}{2\pi T_0^2\tau_0},
\end{equation}
both integrals in Eq.~\eqref{eq:strong-general} are elementary:
\begin{equation}
  \boxed{
  \Sigma(t_f)\simeq\mathcal H(0)
  +\frac{\epsilon}{12}\left[u_0+d\left(1-\frac{T_c}{T_0}\right)\right],
  \qquad \epsilon\to0,}
  \label{eq:strong-bjorken}
\end{equation}
where $T_c=T(t_f)$.  In the opposite limit, $\gamma t_f\ll1$, the distribution
is frozen at its initial form while the reference state cools.  With
$\rho=T_0/T_c$, Eq.~\eqref{eq:cooling-production} integrates to
\begin{equation}
  \boxed{
  \Sigma(t_f)\simeq\frac1\epsilon\left[
  \frac32\left(r_0^2+dx_0\right)(\rho^2-1)
  -6d(\rho-1)
  +\frac{3d}{x_0}\ln\rho\right],
  \qquad \epsilon\to\infty.}
  \label{eq:weak-bjorken}
\end{equation}

\begin{figure}[t]
  \centering
  \includegraphics[width=0.82\linewidth]{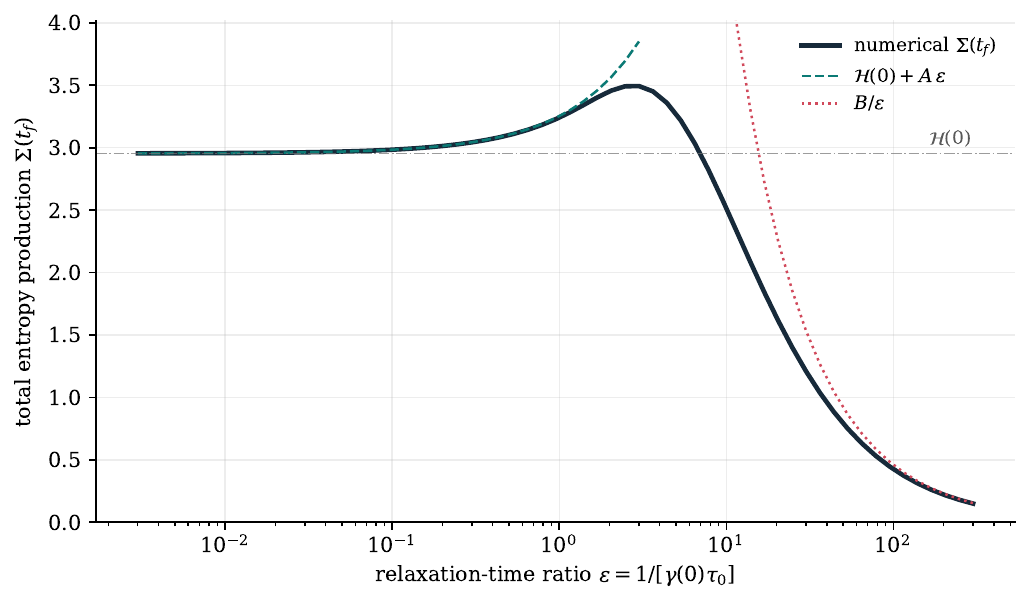}
  \caption{Total production of the driven Gaussian benchmark in the Bjorken
  temperature protocol as a function of the relaxation-time ratio
  $\epsilon=1/[\gamma(0)\tau_0]$, for $d=3$, $x_0=0.25$, $r_0^2=4$, and
  $T_0/T_c=0.40/0.155$.  Dashed: strong-coupling limit,
  Eq.~\eqref{eq:strong-bjorken}.  Dotted: weak-coupling limit,
  Eq.~\eqref{eq:weak-bjorken}.  The dash-dotted line marks $\mathcal H(0)$.}
  \label{fig:gaussian-scan}
\end{figure}

Equations~\eqref{eq:strong-bjorken} and \eqref{eq:weak-bjorken} settle the
dependence on coupling.  If $u_0+d(1-T_c/T_0)>0$, the total production rises
above $\mathcal H(0)$ as the relaxation time increases from zero, whereas it
vanishes as $1/\epsilon$ at weak coupling.  It must therefore reach a maximum,
exceeding $\mathcal H(0)$, at intermediate coupling.  The condition holds for
every initial state that is at least as hot as the bath, $u_0\geq0$, and also
for moderately colder ones.  Figure~\ref{fig:gaussian-scan} compares the
asymptotes with the exact solution for $d=3$, $x_0=0.25$, $r_0^2=4$, and
$T_0/T_c=0.40/0.155$.  The strong-coupling slope is $0.299$, the weak-coupling
coefficient is $46.0$, and the maximum $\Sigma(t_f)=3.49$ occurs at
$\epsilon\approx3$.  The deviation from Eq.~\eqref{eq:strong-bjorken} decreases
as $\epsilon^2$, and the numerical solution satisfies
Eq.~\eqref{eq:sigma-u} to $10^{-12}$.

\subsection{Longitudinal expansion as a separate drive}
\label{sec:expansion}

The temperature profile in Eq.~\eqref{eq:bjorken-temperature} does not by
itself constitute full Bjorken kinematics.  In a boost-invariant local frame,
free streaming also redshifts the longitudinal momentum.  At the present level
this is represented by
\begin{equation}
  \partial_t f=\vect\nabla_p\!\cdot
  [\gamma\vect p f+\gamma MT\vect\nabla_p f]
  +\frac{1}{\tau_0+t}\partial_{p_z}(p_z f).
  \label{eq:bjorken-fp}
\end{equation}
Gaussian closure survives, but the transverse and longitudinal variances
$\theta_T$ and $\theta_L$ evolve differently:
\begin{subequations}
\begin{align}
  \dot m_{x,y}&=-\gamma m_{x,y},
  &\dot m_z&=-\left(\gamma+\frac{1}{\tau_0+t}\right)m_z,\\
  \dot\theta_T&=-2\gamma(\theta_T-MT),
  &\dot\theta_L&=-2\left(\gamma+\frac{1}{\tau_0+t}\right)\theta_L+2\gamma MT.
\end{align}
\label{eq:anisotropic-moments}
\end{subequations}
Only the collision current belongs in the irreversible production,
\begin{equation}
  \Pi_{\rm coll}=\gamma\left[
  \frac{|\vect m|^2}{MT}
  +2\frac{(\theta_T-MT)^2}{MT\theta_T}
  +\frac{(\theta_L-MT)^2}{MT\theta_L}\right].
  \label{eq:anisotropic-production}
\end{equation}
The longitudinal redshift changes the differential entropy at the kinematic
rate $\dot S_{\rm exp}=-1/(\tau_0+t)$.  Consequently the relative-entropy
balance is
\begin{align}
  \dot{\mathcal H}={}&-\Pi_{\rm coll}
  +\frac{\dot T}{2T}\left[
  3-\frac{|\vect m|^2+2\theta_T+\theta_L}{MT}\right]\notag\\
  &+\frac{1}{\tau_0+t}\left[
  1-\frac{m_z^2+\theta_L}{MT}\right].
  \label{eq:bjorken-driven-h}
\end{align}
The last line is the phase-space contraction of the reduced momentum
distribution produced by the expansion, not bath entropy production.  The
numerical benchmarks below deliberately retain only the temperature protocol, so
that the thermalization and cooling terms can be tested without conflating them
with this kinematic drive.

\section{Relativistic detailed-balance extension}
\label{sec:relativistic}

The Gaussian process is a controlled analytic benchmark, but its stationary
distribution is nonrelativistic.  A relativistic heavy quark should approach the
J\"uttner density
\begin{equation}
  f_{\mathrm{eq}}(\vect p)
  =Z^{-1}\exp[-E_p/T],
  \qquad E_p=\sqrt{M^2+|\vect p|^2}.
  \label{eq:juttner}
\end{equation}
Relativistic Langevin processes require care with multiplicative noise and the
choice of stochastic discretization
\cite{HanggiThomas1982,Dunkel2009,Akamatsu2009,He2023}.  The cleanest starting
point is the Fokker--Planck current itself, written in detailed-balance form,
\begin{equation}
  \vect J
  =-\mathsf D(\vect p,T)
  \left[\vect\nabla_p f+f\vect\nabla_p(E_p/T)\right],
  \qquad
  \partial_t f=-\vect\nabla_p\cdot\vect J.
  \label{eq:relativistic-current}
\end{equation}
For an isotropic medium the positive diffusion tensor can be decomposed as
\begin{equation}
  \mathsf D
  =D_T(\vect p,T)
  \left(I-\hat{\vect p}\hat{\vect p}^{\mathsf T}\right)
  +D_L(\vect p,T)
  \hat{\vect p}\hat{\vect p}^{\mathsf T}.
  \label{eq:diffusion-tensor}
\end{equation}
Equation~\eqref{eq:relativistic-current} guarantees
$\vect J[f_{\mathrm{eq}}]=0$.  The drift of an equivalent stochastic
differential equation contains convention-dependent derivative terms when
$\mathsf D$ depends on momentum, so specifying only a drag coefficient without
the discretization prescription is insufficient.

Since
\begin{equation}
  \vect J=-f\mathsf D\vect\nabla_p
  \ln\frac{f}{f_{\mathrm{eq}}},
\end{equation}
the corresponding entropy production is
\begin{equation}
  \boxed{
  \Pi_{\mathrm{rel}}
  =\int \dd^3p\,f
  \left(\vect\nabla_p\ln\frac{f}{f_{\mathrm{eq}}}\right)^{\!\mathsf T}
  \mathsf D
  \left(\vect\nabla_p\ln\frac{f}{f_{\mathrm{eq}}}\right)
  \geq0.}
  \label{eq:relativistic-production}
\end{equation}
For fixed $T$, integration by parts yields
\begin{equation}
  \frac{\dd}{\dd t}\KL(f\Vert f_{\mathrm{eq}})
  =-\Pi_{\mathrm{rel}}.
  \label{eq:relativistic-h-theorem}
\end{equation}
For a prescribed time-dependent reference state, the exact balance is
\begin{equation}
  \frac{\dd}{\dd t}\KL[f\Vert f_{\mathrm{eq}}(t)]
  =-\Pi_{\mathrm{rel}}
  -\int \dd^3p\,f\,\partial_t\ln f_{\mathrm{eq}}.
  \label{eq:relativistic-driven}
\end{equation}
Equations~\eqref{eq:relativistic-production} and
\eqref{eq:relativistic-driven} are the natural relativistic counterparts of
Eqs.~\eqref{eq:production-main} and \eqref{eq:driven-h-theorem}.  They remain
valid for distinct longitudinal and transverse diffusion coefficients but, in
general, cannot be reduced to the first two moments alone.

\subsection{Relativistic cooling benchmark}

For an isotropic distribution we choose the scalar diffusion tensor
\begin{equation}
  \mathsf D=D(p,t)I_3,
  \qquad D(p,t)=\gamma(t)E_pT(t).
  \label{eq:scalar-relativistic-diffusion}
\end{equation}
The radial equation is then
\begin{equation}
  \partial_t f
  =\frac{1}{p^2}\partial_p\left\{p^2D(p,t)
  \left[\partial_p f+\frac{p}{E_pT(t)}f\right]\right\}.
  \label{eq:radial-relativistic-fp}
\end{equation}
This choice gives the drag current $-\gamma\vect p f$ and reduces at zero
momentum to the usual Einstein relation.  We therefore use
\begin{equation}
  \gamma(t)=\frac{2\pi T^2(t)}{c_D M\hbar c},
  \qquad c_D=2\pi T D_s,
  \label{eq:relativistic-gamma}
\end{equation}
where the factor $\hbar c$ converts the rate to fm$^{-1}$.

The time-dependent J\"uttner state makes the driving term in
Eq.~\eqref{eq:relativistic-driven} explicit:
\begin{equation}
  \dot{\mathcal H}=-\Pi_{\rm rel}
  -\frac{\dot T}{T^2}
  \left(\avg{E_p}_f-\avg{E_p}_{\rm eq}\right).
  \label{eq:relativistic-driven-energy}
\end{equation}
This identity is used as a numerical closure test rather than imposed in the
evolution.

We take $T_0=0.40$ GeV, $T_c=0.155$ GeV, $\tau_0=0.60$ fm/$c$,
$M_c=1.5$ GeV, and $M_b=4.8$ GeV.  The evolution ends at
$T(t_f)=T_c$, or $t_f=9.71$ fm/$c$.  The smooth initial distribution is a
J\"uttner density at $T_{Q,0}=0.80$ GeV.  It is a thermal-like benchmark, not
a fit to a perturbative production spectrum, but it remains well defined for
charm where a Maxwell initialization would not be controlled.

Equation~\eqref{eq:radial-relativistic-fp} is solved on a conservative radial
grid with 280 cells over $0<p<14$ GeV.  A Scharfetter--Gummel interface flux
preserves the discrete J\"uttner state, and no-flux conditions are imposed at
both boundaries.  The time integration uses an implicit variable-step method.
The entropy production is evaluated from the same interface current, so the
semi-discrete entropy identity is exact up to time integration.  Increasing the
grid to 420 cells changes the reported totals by less than one part in
$10^3$.  The diffusion scan below uses 150 cells and 1080 output times, and its
balance residual remains below $6\times10^{-5}$ for every value of $c_D$.

\begin{table}[t]
\centering
\caption{Relativistic cooling benchmark for $c_D=4$.  The last column is the
maximum absolute residual of Eq.~\eqref{eq:relativistic-driven-energy}.}
\label{tab:relativistic-cooling}
\begin{tabular}{lcccccc}
\toprule
Species & $M$ [GeV] & $\gamma(0)$ [fm$^{-1}$] & $\Sigma(t_f)$
& $\mathcal H(0)$ & $\mathcal H(t_f)$ & $\max|R|$ \\
\midrule
Charm  & 1.5 & 0.849 & 2.221 & 0.738 & 0.101 & $5.9\times10^{-6}$ \\
Bottom & 4.8 & 0.265 & 1.964 & 0.587 & 1.028 & $1.6\times10^{-5}$ \\
\bottomrule
\end{tabular}
\end{table}

\begin{figure}[t]
  \centering
  \includegraphics[width=\linewidth]{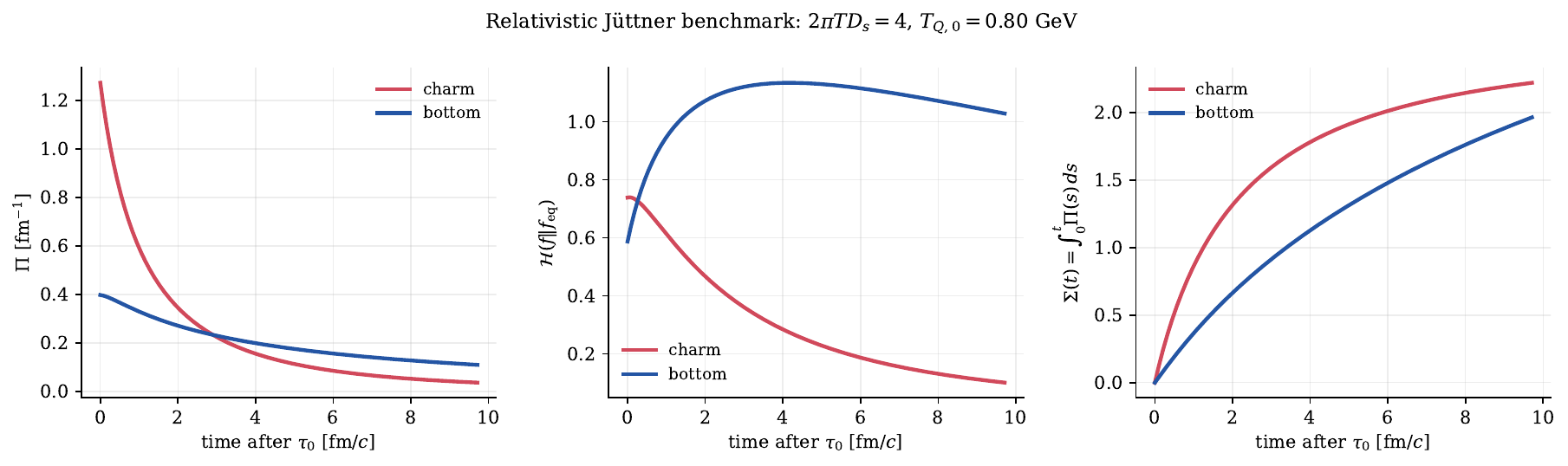}
  \caption{Relativistic charm and bottom evolution for $c_D=4$ in the
  temperature-only Bjorken protocol.  From left to right: instantaneous entropy
  production, relative entropy to the instantaneous J\"uttner state, and
  accumulated production.}
  \label{fig:relativistic-cooling}
\end{figure}

Figure~\ref{fig:relativistic-cooling} shows the rate, relative entropy, and
accumulated production.  Charm responds rapidly enough that its final distance
from the instantaneous target is small.  Bottom retains a larger relative
entropy even though its integrated production is substantial.  In both cases
$\Sigma(t_f)$ exceeds the initial relative entropy because cooling continually
displaces the target, exactly as accounted for by
Eq.~\eqref{eq:relativistic-driven-energy}.

\subsection{Coupling dependence and linear-response limits}
\label{sec:relativistic-limits}

The Gaussian limits of Sec.~\ref{sec:limits} generalize directly.  Write the
evolution as $\partial_tf=\gamma(t)\mathcal L_Tf$, where $\mathcal L_T$ is the
generator of Eq.~\eqref{eq:radial-relativistic-fp} at unit drag.  It is
negative semidefinite in the inner product weighted by $f_{\mathrm{eq}}^{-1}$
and annihilates only $f_{\mathrm{eq}}$, so $\mathcal L_T^{-1}$ is well defined
on functions with zero integral.  At strong coupling, the transient
$f-f_{\mathrm{eq}}=\ee^{\gamma\mathcal L_T t}(f_0-f_{\mathrm{eq}})$ and the
quasistatic lag
$f-f_{\mathrm{eq}}\simeq\gamma^{-1}\mathcal L_T^{-1}\partial_tf_{\mathrm{eq}}$,
inserted in Eq.~\eqref{eq:relativistic-driven-energy}, give
\begin{equation}
  \Sigma(t_f)\simeq\mathcal H(0)
  +\frac{|\dot T(0)|}{\gamma(0)T_0^2}
  \avg{E_p,-\mathcal L_{T_0}^{-1}(f_0-f_{\mathrm{eq}})}
  +\int_0^{t_f}\frac{\dot T^2}{\gamma}\,\zeta(T)\,\dd t,
  \label{eq:relativistic-strong}
\end{equation}
with $\avg{E_p,g}=\int\dd^3p\,E_pg$ and the friction coefficient
\begin{equation}
  \zeta(T)=\frac{1}{T^4}
  \avg{E_p,-\mathcal L_T^{-1}\left[(E_p-\avg{E_p}_{\rm eq})f_{\mathrm{eq}}\right]}
  \geq0.
  \label{eq:relativistic-friction}
\end{equation}
For the Ornstein--Uhlenbeck generator the energy relaxes at the single rate
$2\gamma$, and Eqs.~\eqref{eq:relativistic-strong} and
\eqref{eq:relativistic-friction} reduce to Eq.~\eqref{eq:strong-general}.
Because $\gamma\propto1/c_D$, Eq.~\eqref{eq:relativistic-strong} has the form
$\Sigma\simeq\mathcal H(0)+a\,c_D$.  At weak coupling the distribution stays
frozen at $f_0$, and $\Sigma\simeq b/c_D$, where $b/c_D=\int\gamma\,\Pi_1\,\dd t$
and $\Pi_1$ is the production of $f_0$ at unit drag relative to the
instantaneous J\"uttner state.  We evaluate $a$ and $b$ with the same discrete
generator used in the evolution.  Table~\ref{tab:relativistic-limits} lists the
coefficients.

\begin{table}[t]
\centering
\caption{Linear-response coefficients of the relativistic benchmark,
$\Sigma(t_f)\simeq\mathcal H(0)+a\,c_D$ for $c_D\to0$ and
$\Sigma(t_f)\simeq b/c_D$ for $c_D\to\infty$.  The slope $a$ is split into the
transient and quasistatic tracking terms of
Eq.~\eqref{eq:relativistic-strong}.}
\label{tab:relativistic-limits}
\begin{tabular}{lccccc}
\toprule
Species & $\mathcal H(0)$ & $a_{\rm transient}$ & $a_{\rm tracking}$ & $a$ & $b$ \\
\midrule
Charm  & 0.738 & 0.296 & 0.138 & 0.433 & 54.5 \\
Bottom & 0.587 & 0.605 & 0.309 & 0.914 & 17.0 \\
\bottomrule
\end{tabular}
\end{table}

\begin{figure}[t]
  \centering
  \includegraphics[width=0.82\linewidth]{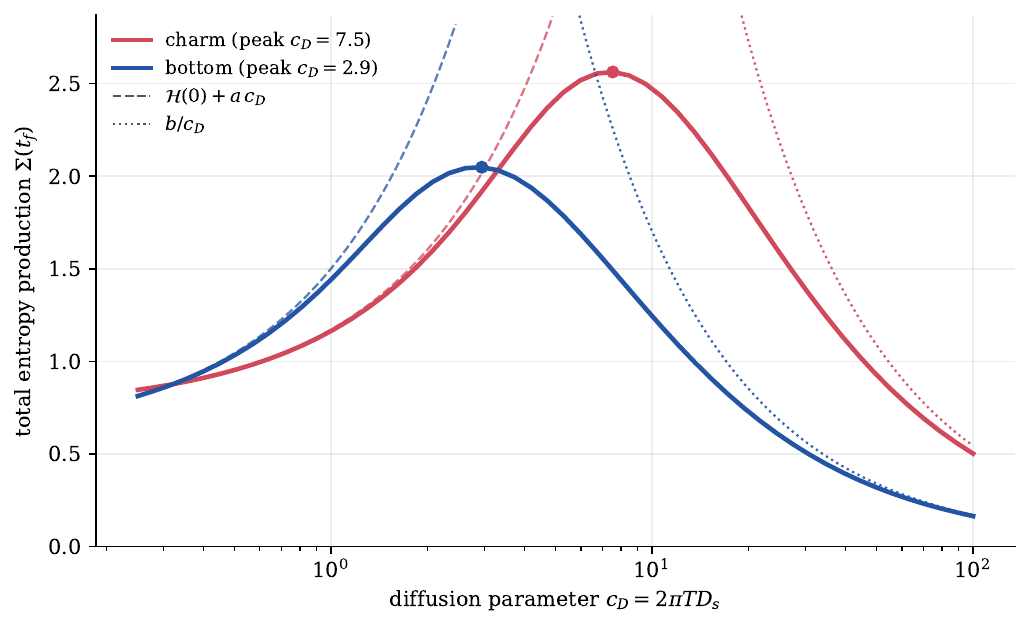}
  \caption{Total relativistic entropy production at $t_f$ over
  $0.25\leq c_D\leq100$.  Dashed curves show the strong-coupling limit
  $\mathcal H(0)+a\,c_D$ and dotted curves the weak-coupling limit $b/c_D$,
  with the coefficients of Table~\ref{tab:relativistic-limits}.  Both charm and bottom
  have an interior maximum.}
  \label{fig:relativistic-scan}
\end{figure}

The full diffusion scan in Fig.~\ref{fig:relativistic-scan} confirms these
limits and resolves the maximum for both species.  At $c_D=0.25$ the
strong-coupling asymptote differs from the numerical total by $0.04\%$ for charm
and $0.3\%$ for bottom.  At $c_D=100$ the weak-coupling asymptote is approached
from above, within $8\%$ for charm and $3\%$ for bottom.  Because $a>0$ for
both species, the argument of Sec.~\ref{sec:limits} applies unchanged: the
production rises above $\mathcal H(0)$, then falls as $1/c_D$, and has an
interior maximum.  Charm reaches it near $c_D=7.5$ and bottom near $c_D=2.9$.
The transient term dominates $a$ for both species.  In this benchmark the lag of
the initial relaxation behind the cooling bath is therefore a larger source of
excess dissipation than quasistatic tracking.

\section{Physical interpretation and use}
\label{sec:limitations}

The calculation isolates the momentum-space irreversibility of a dilute heavy
probe.  This is the level at which a Langevin transport model is defined, so the
reduction is a feature of the observable rather than an inconsistency.  Spatial
propagation, local fluid velocity, radiative loss, and hadronization are needed
to predict $R_{AA}$, $v_2$, or reconstructed heavy-flavor yields
\cite{Prino2016,Cao2013,Ke2018,He2023}, but they do not alter the entropy identity
for the momentum-space Fokker--Planck current.  In a realistic simulation they
would determine the trajectory, the local temperature, and the transport
coefficients inserted into that identity.

Gaussian closure is exact for Eq.~\eqref{eq:langevin} when the initial state is
Gaussian.  A measured or perturbatively generated heavy-quark spectrum need not
have that form, so its first two moments do not determine the full production.
The current formula Eq.~\eqref{eq:production-current} remains valid for the
complete distribution, and the Gaussian solution provides a stringent analytic
test for numerical Langevin codes.  The physical calculation in
Sec.~\ref{sec:relativistic} does not use the Gaussian or nonrelativistic
approximation.

The Markovian assumption has a similarly precise role.  Boltzmann comparisons
show that soft collisions and a large mass-to-temperature ratio favor the
Langevin limit \cite{Das2014}, whereas finite correlation times can modify the
approach to equilibrium \cite{Ruggieri2023}.  These observations set the range
over which the benchmark is expected to be quantitative, but they do not
weaken the entropy balance within that range.  A related assumption is that
the fluctuation-dissipation relation holds instantaneously at the local
temperature while the medium evolves.  When the environment is itself driven
by a time-dependent field, the fluctuation-dissipation relation acquires
explicit drive-dependent terms and becomes non-Markovian even if the undriven
particle-bath system is Markovian, as shown for a quantum Langevin model in
Ref.~\cite{GambaCuiZaccone2025}.  Whether corrections of this kind matter for
heavy quarks in a rapidly cooling plasma lies outside the present treatment.
If memory effects are retained, the probability space must be enlarged or a
non-Markovian entropy balance must be used.

The temperature-only Bjorken profile and constant values of $2\pi T D_s$ expose
the mechanism with few moving parts.  Modern calculations find temperature,
momentum, and pre-equilibrium dependence in the diffusion tensor
\cite{Boguslavski2024,Du2024,He2023}.  These refinements change the
coefficients $a$ and $b$ and therefore the location and height of the maximum,
but not the positive quadratic form in Eq.~\eqref{eq:relativistic-production}
or the linear-response structure of Eq.~\eqref{eq:relativistic-strong}.
Equation~\eqref{eq:bjorken-driven-h} shows separately how the balance is
extended when longitudinal redshift is included.

Finally, $S$ and $\Pi$ refer to the coarse-grained heavy-quark momentum sector.
They should not be added to a hydrodynamic entropy budget without a consistent
open-system accounting of the bath.  Their value is instead diagnostic: they
measure the irreversible loss of information carried by the heavy-quark
momentum distribution under a specified medium protocol.

\section{Conclusions}
\label{sec:conclusions}

The heavy-quark Langevin equation carries a well-defined and useful entropy
balance.  For a Gaussian momentum distribution, its irreversible production
splits exactly into relaxation of the mean and relaxation of the width.  Both
pieces are non-negative, and in a static bath their integral is fixed by the
initial relative entropy.  A delta-function state lies outside this
differential-entropy description, so practical evaluations require a resolved
initial spectrum.

Cooling changes the problem in an essential way.  The quark relaxes toward a
target that is itself moving, so accumulated production is controlled by both
the interaction rate and the nonequilibrium lag.  For the Bjorken temperature
protocol, the total production exceeds the initial relative entropy at strong
coupling by a term linear in the relaxation time.  That term combines the
thermodynamic-length dissipation of quasistatic tracking with the cost of an
initial relaxation that proceeds against a moving target.  At weak coupling the
production vanishes inversely with the relaxation time.  The two limits together
guarantee maximal dissipation at intermediate coupling.  Longitudinal expansion
adds a further kinematic drive that does not belong to the collision production.

The relativistic calculation puts this structure on a physical footing.  For
both charm and bottom, linear-response coefficients computed from the static
generator reproduce the numerical limits, and the total production has its
maximum at intermediate $2\pi T D_s$.  The same functional can be evaluated
locally in a hydrodynamic Langevin simulation with momentum-dependent transport
coefficients, providing a quantitative measure of where and how efficiently
heavy flavor loses memory of its initial state.

\section*{CRediT authorship contribution statement}

\textbf{A. M. Silva:} Conceptualization, Methodology, Software, Formal
analysis, Validation, Visualization, Writing -- original draft, Writing --
review \& editing.

\section*{Declaration of competing interest}

The author declares that there are no known competing financial interests or
personal relationships that could have appeared to influence the work reported
in this paper.

\section*{Funding}

This research did not receive any specific grant from funding agencies in the
public, commercial, or not-for-profit sectors.

\section*{Declaration of generative AI and AI-assisted technologies in the writing process}

During the preparation of this work the author used Claude (Anthropic) to check
derivations, run and extend numerical verification, and revise the text and the
figure-generation code.  After using this tool, the author reviewed and edited
the content as needed and takes full responsibility for the content of the
publication.

\section*{Data availability}

The scripts that generate all figures and tables, and the tabulated numerical
results, are provided as supplementary material.  The calculations require
NumPy, SciPy, and Matplotlib.

\bibliographystyle{elsarticle-num}
\bibliography{heavy_quark_entropy}

@article{Landi2013,
  author = {Landi, Gabriel T. and Tom{\'e}, T{\^a}nia and de Oliveira, M{\'a}rio J.},
  title = {Entropy production in linear {L}angevin systems},
  journal = {Journal of Physics A: Mathematical and Theoretical},
  volume = {46},
  number = {39},
  pages = {395001},
  year = {2013},
  doi = {10.1088/1751-8113/46/39/395001}
}

@article{Seifert2012,
  author = {Seifert, Udo},
  title = {Stochastic thermodynamics, fluctuation theorems and molecular machines},
  journal = {Reports on Progress in Physics},
  volume = {75},
  number = {12},
  pages = {126001},
  year = {2012},
  doi = {10.1088/0034-4885/75/12/126001}
}

@article{Dunkel2009,
  author = {Dunkel, J{\"o}rn and H{\"a}nggi, Peter},
  title = {Relativistic {B}rownian motion},
  journal = {Physics Reports},
  volume = {471},
  number = {1},
  pages = {1--73},
  year = {2009},
  doi = {10.1016/j.physrep.2008.12.001}
}

@article{Svetitsky1988,
  author = {Svetitsky, Benjamin},
  title = {Diffusion of charmed quarks in the quark-gluon plasma},
  journal = {Physical Review D},
  volume = {37},
  number = {9},
  pages = {2484--2491},
  year = {1988},
  doi = {10.1103/PhysRevD.37.2484}
}

@article{Moore2005,
  author = {Moore, Guy D. and Teaney, Derek},
  title = {How much do heavy quarks thermalize in a heavy ion collision?},
  journal = {Physical Review C},
  volume = {71},
  number = {6},
  pages = {064904},
  year = {2005},
  doi = {10.1103/PhysRevC.71.064904}
}

@article{vanHees2005,
  author = {van Hees, Hendrik and Rapp, Ralf},
  title = {Thermalization of heavy quarks in the quark-gluon plasma},
  journal = {Physical Review C},
  volume = {71},
  number = {3},
  pages = {034907},
  year = {2005},
  doi = {10.1103/PhysRevC.71.034907}
}

@article{Akamatsu2009,
  author = {Akamatsu, Yukinao and Hatsuda, Tetsuo and Hirano, Tetsufumi},
  title = {Heavy quark diffusion with relativistic {L}angevin dynamics in the quark-gluon fluid},
  journal = {Physical Review C},
  volume = {79},
  number = {5},
  pages = {054907},
  year = {2009},
  doi = {10.1103/PhysRevC.79.054907}
}

@article{Beraudo2009,
  author = {Beraudo, Andrea and De Pace, Arturo and Alberico, Wanda M. and Molinari, Andrea},
  title = {Transport properties and {L}angevin dynamics of heavy quarks and quarkonia in the quark-gluon plasma},
  journal = {Nuclear Physics A},
  volume = {831},
  number = {1--2},
  pages = {59--90},
  year = {2009},
  doi = {10.1016/j.nuclphysa.2009.09.002}
}

@article{He2023,
  author = {He, Min and van Hees, Hendrik and Rapp, Ralf},
  title = {Heavy-quark diffusion in the quark-gluon plasma},
  journal = {Progress in Particle and Nuclear Physics},
  volume = {130},
  pages = {104020},
  year = {2023},
  doi = {10.1016/j.ppnp.2023.104020}
}

@article{Prino2016,
  author = {Prino, Francesco and Rapp, Ralf},
  title = {Open heavy flavor in {QCD} matter and in nuclear collisions},
  journal = {Journal of Physics G: Nuclear and Particle Physics},
  volume = {43},
  number = {9},
  pages = {093002},
  year = {2016},
  doi = {10.1088/0954-3899/43/9/093002}
}

@article{Bjorken1983,
  author = {Bjorken, J. D.},
  title = {Highly relativistic nucleus-nucleus collisions: The central rapidity region},
  journal = {Physical Review D},
  volume = {27},
  number = {1},
  pages = {140--151},
  year = {1983},
  doi = {10.1103/PhysRevD.27.140}
}

@article{Du2024,
  author = {Du, Xiaojian},
  title = {Heavy quark drag and diffusion coefficients in the prehydrodynamic {QCD} plasma},
  journal = {Physical Review C},
  volume = {109},
  number = {1},
  pages = {014901},
  year = {2024},
  doi = {10.1103/PhysRevC.109.014901}
}

@article{Ke2018,
  author = {Ke, Weiyao and Xu, Yingru and Bass, Steffen A.},
  title = {Linearized {B}oltzmann-{L}angevin model for heavy quark transport in hot and dense {QCD} matter},
  journal = {Physical Review C},
  volume = {98},
  number = {6},
  pages = {064901},
  year = {2018},
  doi = {10.1103/PhysRevC.98.064901}
}

@article{Ruggieri2023,
  author = {Ruggieri, Marco and Das, Santosh K.},
  title = {Thermalization and isotropization of heavy quarks in a non-{M}arkovian medium in high-energy nuclear collisions},
  journal = {Physical Review D},
  volume = {108},
  number = {5},
  pages = {054026},
  year = {2023},
  doi = {10.1103/PhysRevD.108.054026}
}

@book{Risken1989,
  author = {Risken, Hannes},
  title = {The {F}okker--{P}lanck Equation: Methods of Solution and Applications},
  edition = {2},
  publisher = {Springer},
  address = {Berlin},
  year = {1989},
  doi = {10.1007/978-3-642-61544-3}
}

@book{Sekimoto2010,
  author = {Sekimoto, Ken},
  title = {Stochastic Energetics},
  series = {Lecture Notes in Physics},
  volume = {799},
  publisher = {Springer},
  address = {Berlin},
  year = {2010},
  doi = {10.1007/978-3-642-05411-2}
}

@article{HanggiThomas1982,
  author = {H{\"a}nggi, Peter and Thomas, Harry},
  title = {Stochastic processes: Time evolution, symmetries and linear response},
  journal = {Physics Reports},
  volume = {88},
  number = {4},
  pages = {207--319},
  year = {1982},
  doi = {10.1016/0370-1573(82)90045-X}
}

@article{SpinneyFord2012,
  author = {Spinney, Richard E. and Ford, Ian J.},
  title = {Entropy production in full phase space for continuous stochastic dynamics},
  journal = {Physical Review E},
  volume = {85},
  number = {5},
  pages = {051113},
  year = {2012},
  doi = {10.1103/PhysRevE.85.051113}
}

@article{SchmiedlSeifert2008,
  author = {Schmiedl, Tim and Seifert, Udo},
  title = {Efficiency at maximum power: An analytically solvable model for stochastic heat engines},
  journal = {Europhysics Letters},
  volume = {81},
  number = {2},
  pages = {20003},
  year = {2008},
  doi = {10.1209/0295-5075/81/20003}
}

@article{FordLakerCharlesworth2015,
  author = {Ford, Ian J. and Laker, Zachary P. L. and Charlesworth, Henry J.},
  title = {Stochastic entropy production arising from nonstationary thermal transport},
  journal = {Physical Review E},
  volume = {92},
  number = {4},
  pages = {042108},
  year = {2015},
  doi = {10.1103/PhysRevE.92.042108}
}

@article{SpinneyFord2012PRL,
  author = {Spinney, Richard E. and Ford, Ian J.},
  title = {Nonequilibrium thermodynamics of stochastic systems with odd and even variables},
  journal = {Physical Review Letters},
  volume = {108},
  number = {17},
  pages = {170603},
  year = {2012},
  doi = {10.1103/PhysRevLett.108.170603}
}

@article{BlaizotEscobedo2018,
  author = {Blaizot, Jean-Paul and Escobedo, Miguel Angel},
  title = {Approach to equilibrium of a quarkonium in a quark-gluon plasma},
  journal = {Physical Review D},
  volume = {98},
  number = {7},
  pages = {074007},
  year = {2018},
  doi = {10.1103/PhysRevD.98.074007}
}

@article{HeFriesRapp2012,
  author = {He, Min and Fries, Rainer J. and Rapp, Ralf},
  title = {Heavy-quark diffusion and hadronization in quark-gluon plasma},
  journal = {Physical Review C},
  volume = {86},
  number = {1},
  pages = {014903},
  year = {2012},
  doi = {10.1103/PhysRevC.86.014903}
}

@article{Cao2013,
  author = {Cao, Shanshan and Qin, Guang-You and Bass, Steffen A.},
  title = {Heavy quark dynamics and hadronization in ultra-relativistic heavy-ion collisions: Collisional versus radiative energy loss},
  journal = {Physical Review C},
  volume = {88},
  number = {4},
  pages = {044907},
  year = {2013},
  doi = {10.1103/PhysRevC.88.044907}
}

@article{Das2014,
  author = {Das, Santosh K. and Scardina, Francesco and Plumari, Salvatore and Greco, Vincenzo},
  title = {Heavy-flavor in-medium momentum evolution: {L}angevin versus {B}oltzmann approach},
  journal = {Physical Review C},
  volume = {90},
  number = {4},
  pages = {044901},
  year = {2014},
  doi = {10.1103/PhysRevC.90.044901}
}

@article{Boguslavski2024,
  author = {Boguslavski, Kirill and Kurkela, Aleksi and Lappi, Tuomas and Lindenbauer, Florian and Peuron, Jarkko},
  title = {Heavy quark diffusion coefficient in heavy-ion collisions via kinetic theory},
  journal = {Physical Review D},
  volume = {109},
  number = {1},
  pages = {014025},
  year = {2024},
  doi = {10.1103/PhysRevD.109.014025}
}

@article{SalamonBerry1983,
  author = {Salamon, Peter and Berry, R. Stephen},
  title = {Thermodynamic length and dissipated availability},
  journal = {Physical Review Letters},
  volume = {51},
  number = {13},
  pages = {1127--1130},
  year = {1983},
  doi = {10.1103/PhysRevLett.51.1127}
}

@article{SivakCrooks2012,
  author = {Sivak, David A. and Crooks, Gavin E.},
  title = {Thermodynamic metrics and optimal paths},
  journal = {Physical Review Letters},
  volume = {108},
  number = {19},
  pages = {190602},
  year = {2012},
  doi = {10.1103/PhysRevLett.108.190602}
}

@article{GambaCuiZaccone2025,
  author = {Gamba, Daniele and Cui, Bingyu and Zaccone, Alessio},
  title = {Open quantum systems with particle and bath driven by time-dependent fields},
  journal = {Physical Review A},
  volume = {112},
  number = {1},
  pages = {012207},
  year = {2025},
  doi = {10.1103/w3vk-wx62}
}

\end{document}